\newif\ifmnras
\mnrastrue
\ifmnras
\documentclass[useAMS,usenatbib,usedcolumn]{mnras}
\else
\documentclass[apj]{emulateapj}
\fi

\usepackage{graphics,epsf}
\usepackage{amsmath}                
\usepackage{amsfonts}               
\usepackage{amssymb}                
\usepackage{epsfig}                 
\usepackage{graphicx}
\usepackage{float}

\newcommand{\cm}{{~\rm cm}}
\newcommand{\km}{{~\rm km}}
\newcommand{\s}{{~\rm s}}

\newcommand{\erg}{{~\rm erg}}
\newcommand{\yr}{{~\rm yr}}

\ifmnras
\else
\newcommand{\nar}{{~\rm New Astronomy Reviews}}
\newcommand{\na}{{~\rm New Astronomy}}
\fi

\ifmnras
\title[iPTF14hls as a common envelope jets supernova]{
Explaining iPTF14hls as a common envelope jets supernova}
\author[N. Soker, A. Gilkis]{Noam Soker$^{1}$\thanks{Contact e-mail: \href{soker@physics.technion.ac.il}{soker@physics.technion.ac.il}}, Avishai Gilkis$^{2}$\thanks{Contact e-mail: \href{agilkis@ast.cam.ac.uk}{agilkis@ast.cam.ac.uk}}
\\
$^{1}$ Department of Physics, Technion -- Israel Institute of Technology, Haifa 3200003, Israel \\
$^{2}$ Institute of Astronomy, University of Cambridge, Madingley Rise, Cambridge, CB3 0HA, UK 
}
\fi

\begin{document}

\ifmnras
\pagerange{\pageref{firstpage}--\pageref{lastpage}} \pubyear{2017}

\maketitle
\else
\title[iPTF14hls as a common envelope jets supernova]{Explaining iPTF14hls as a common envelope jets supernova}
\author{Noam Soker\altaffilmark{1} \& Avishai Gilkis\altaffilmark{2}}
\altaffiltext{1}{Department of Physics, Technion -- Israel Institute of Technology, Haifa 3200003, Israel; 	soker@physics.technion.ac.il}
\altaffiltext{2}{Institute of Astronomy, University of Cambridge, Madingley Rise, Cambridge, CB3 0HA, UK; agilkis@ast.cam.ac.uk}
\fi

\label{firstpage}

\begin{abstract}
We propose a \textit{common envelope jets supernova} scenario for the enigmatic supernova iPTF14hls where a neutron star that spirals-in inside the envelope of a massive giant star accretes mass and launches jets that power the ejection of the circumstellar shell and a few weeks later the explosion itself. To account for the kinetic energy of the circumstellar gas and the explosion, the neutron star should accrete a mass of  $\approx 0.3 M_\odot$. The tens$\times M_\odot$ of circumstellar gas that accounts for some absorption lines is ejected while the neutron star orbits for about one to several weeks inside the envelope of the giant star. In the last hours of the interaction the neutron star merges with the core, accretes mass, and launches jets that eject the core and the inner envelope to form the explosion itself and the medium where the supernova photosphere resides. The remaining neutron star accretes fallback gas and further powers the supernova. We attribute the 1954 pre-explosion outburst to an eccentric orbit and temporary mass accretion by the neutron star at periastron passage prior to the onset of the common envelope phase.
\ifmnras
\else
\fi
\end{abstract}

\ifmnras
\fi

\section{INTRODUCTION}
\label{sec:intro}

\cite{Arcavietal2017} report the discovery and evolution of the enigmatic type II-P supernova (SN) iPTF14hls (AT 2016bse; Gaia16aog). iPTF14hls has several unusual properties, from which we list here three. (1) The evolution along the light curve is an order of magnitude slower than that of typical type II-P SNe. (2) The light curve has at least five peaks. (3) The absorbing gas is in a circumstellar matter (CSM) that has a fast outflow velocity of $v_{\rm CSM} \approx 6000 \km \s^{-1}$, and with an estimated kinetic energy of $E_{\rm CSM} \approx 10^{52} \erg$ \citep{Arcavietal2017}.  
 
Ejection of mass tens of years to few days before explosion in pre-explosion outbursts was deduced for other SNe, (e.g., \citealt{Foleyetal2007, Mauerhanetal2013, Ofeketal2013, SvirskiNakar2014,  Moriya2015, Goranskijetal2016, Marguttietal2017, Tartagliaetal2016, BoianGroh2017, Liuetal2017, Nyholmetal2017, Yaronetal2017, Pastorelloetal2018}). In these cases the supernova ejecta is thought to collide with the CSM and convert kinetic energy to radiation. Some of the pre-explosion outbursts are observed to be non-spherical (e.g., \citealt{Reillyetal2017} for SN~2009ip). In contrast to these cases, in iPFT14hls there is no evidence for ejecta-CSM collision. 

One of the employed mechanisms to trigger these pre-explosion outbursts involves energy that is carried by waves from the vigorous convection in the core and is deposited in the envelope \citep{QuataertShiode2012, ShiodeQuataert2014}. Magnetic activity might also carry energy from the core to the envelope \citep{SokerGilkis2017}. However, these mechanisms cannot carry an amount of energy even close to $10^{52} \erg$, and they are likely to cause mainly envelope expansion rather than mass ejection (e.g., \citealt{Soker2013, Fuller2017}). Any other mechanism that deposits $\approx 10^{52}\erg$ inside or close to the core will eject the entire hydrogen-rich envelope, and is not applicable to iPTF14hls which is a type II-P SN. 

Since the deposition of energy in the envelope of the commonly observed pre-explosion outbursts seems to be unable to expel much mass, we adopt the view that a companion accreteing mass from the inflated envelope and launching jets supplies most of the energy to eject the mass and power the outburst (e.g., \citealt{KashiSoker2010, Soker2013, McleySoker2014}). The common envelope process where a companion that orbits inside the envelope or inside the core of a giant star accretes mass and launches jets has been discussed before (e.g., \citealt{FryerWoosley1998, ArmitageLivio2000, Soker2004, Chevalier2012}). \cite{FryerWoosley1998}, for example, propose a gamma ray burst model where a black hole that spirals-in all the way to the helium core of a giant accretes mass and launches jets. 
 
We propose a scenario to account for the enigmatic type II-P SN iPTF14hls and its pre-explosion mass ejection that is based on the jet feedback mechanism (for review see \citealt{Soker2016Rev}) that can both facilitate the removal of a common envelope (e.g., \citealt{ArmitageLivio2000, Soker2004, Chevalier2012, ShiberSoker2018})  and power the explosion of most (or all) core collapse supernovae (e.g., \citealt{PapishSoker2011}). In section \ref {sec:removal} we discuss the jet-driven envelope removal of part of the envelope, and in section \ref{sec:explosion} we discuss the much faster removal of the core by jets, i.e., the explosion. We summarize in section \ref{sec:summary}.

\section{PRE-EXPLOSION ENVELOPE EJECTION}
\label{sec:removal}
 
\cite{Arcavietal2017} note that the evolution of some spectral absorption lines is best explained by coming from pre-explosion ejected gas. However, a velocity of $v_{\rm pre} \simeq 4000-8000 \km \s^{-1}$ is not typical for red giant winds. A much more violent event must take place. We take it to be the launching of jets by a companion.
 
\cite{Arcavietal2017} constrain the pre-explosion luminosity to be below $\approx 2 \times 10^{41} \erg$. This upper bound on the luminosity and the kinetic energy of the pre-explosion ejecta of $E_{\rm CSM} \approx 10^{52} \erg$ \citep{Arcavietal2017}, constrain the efficiency of converting energy to optical radiation to be $\la 10^{-3} (\tau_{\rm CE}/1\yr)$, where $\tau_{\rm CE}$ is the duration of the pre-explosion common envelope phase. This is a very low efficiency. The ejection of tens of solar masses at a velocity of thousands of $\km \s^{-1}$ should be like a supernova explosion and lead to a detection if observed.

Our proposed solution is to consider a very short pre-explosion common envelope phase, during the time the SN was not observed and the upper bound on the luminosity does not apply. Before entering the envelope and accreting at a very high rate that enables neutrino cooling, the neutron star accreted at its Eddington limit, which is much below the detection limit. We propose that the neutron star entered the envelope within less than 100 days before discovery. The neutron star cannot bring the envelope to synchronization, and very rapidly, within a few dynamical times or one Keplerian orbital time (e.g., \citealt{Passyetal2012, IvanovaNandez2016}), spirals-in all the way to the core. The spiral-in time is then
\begin{equation}
\tau_{\rm CE} \approx \tau_{\rm Kep} \simeq 15 
\left( \frac{M}{60 M_\odot} \right)^{-1/2}
\left( \frac{R_1}{100 R_\odot} \right)^{3/2} {\rm ~day} ,
     \label{eq:tauCE}
\end{equation}
where $M$ is the total mass of the binary system, and $R_1$ is the radius of the giant star. As the spiraling in might be several times the Keplerian time on the surface, we consider the radius of $R_1=100$ at the onset of the common envelope phase to be an upper limit. Namely, we expect $R_1 \simeq 50-100 R_\odot$ at the onset of the common envelope phase. 

\begin{figure}
\begin{center}
\includegraphics[scale=0.6]{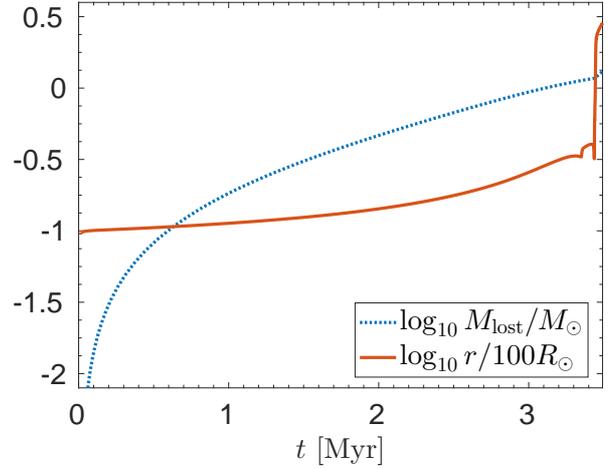}
\vskip +0.5 cm
\caption{Total mass lost and radius of the star as function of time, for a star with an initial mass of $M_\mathrm{ZAMS}=80M_\odot$.}
\label{fig:starhistory}
\end{center}
\end{figure}

\begin{figure}
\begin{center}
\begin{tabular}{c}
\includegraphics[scale=0.58]{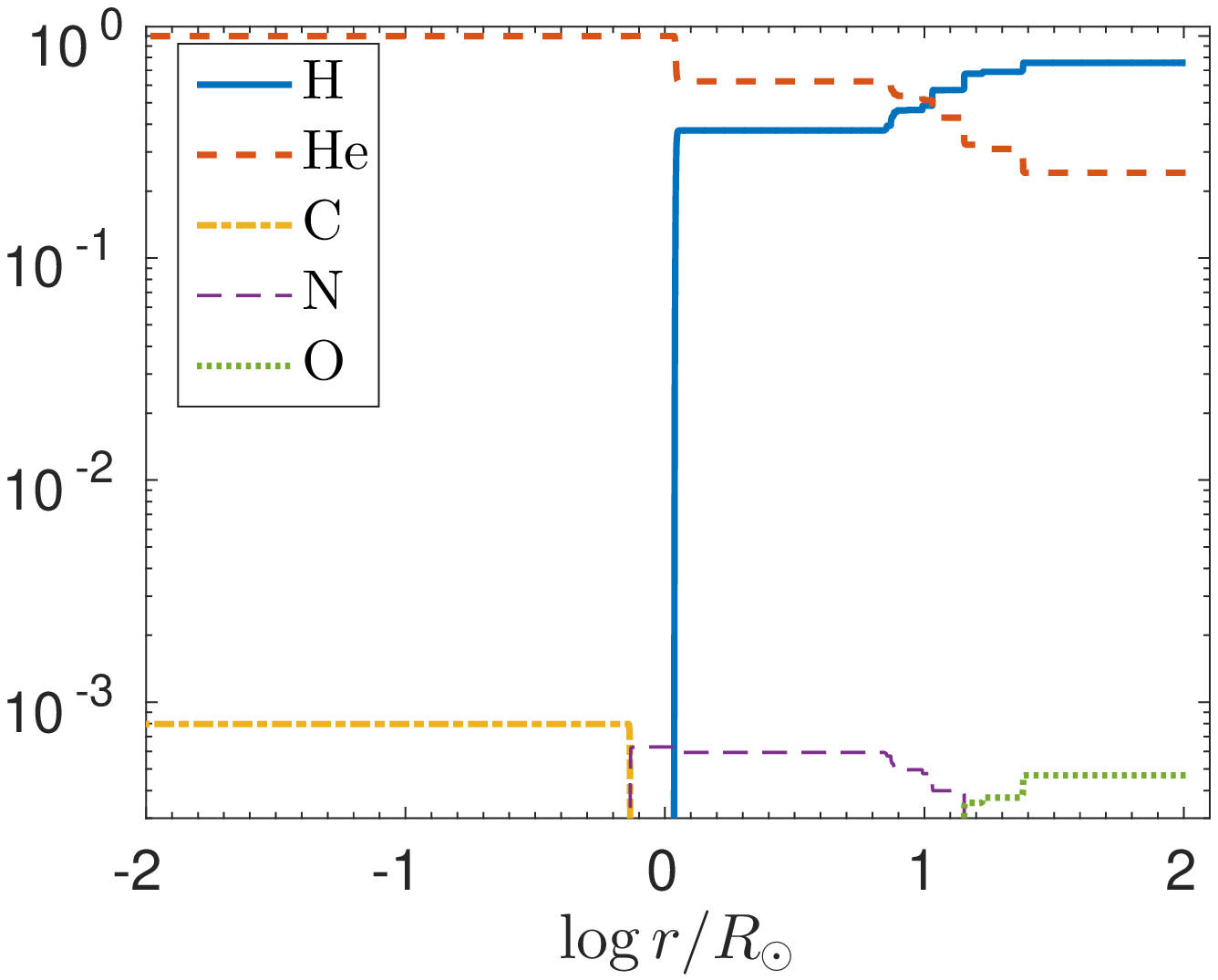}\\
\includegraphics[scale=0.58]{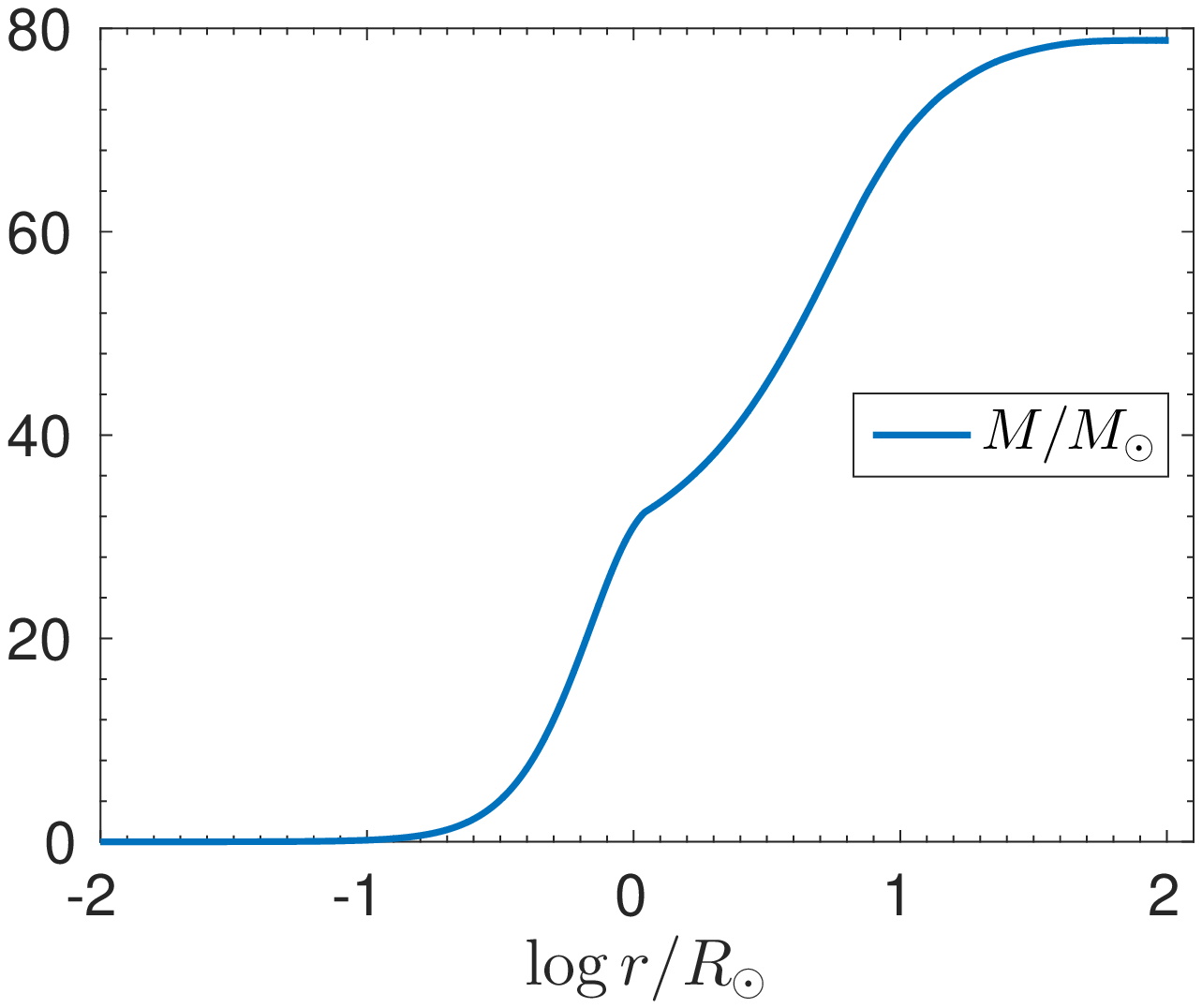}\\
\includegraphics[scale=0.58]{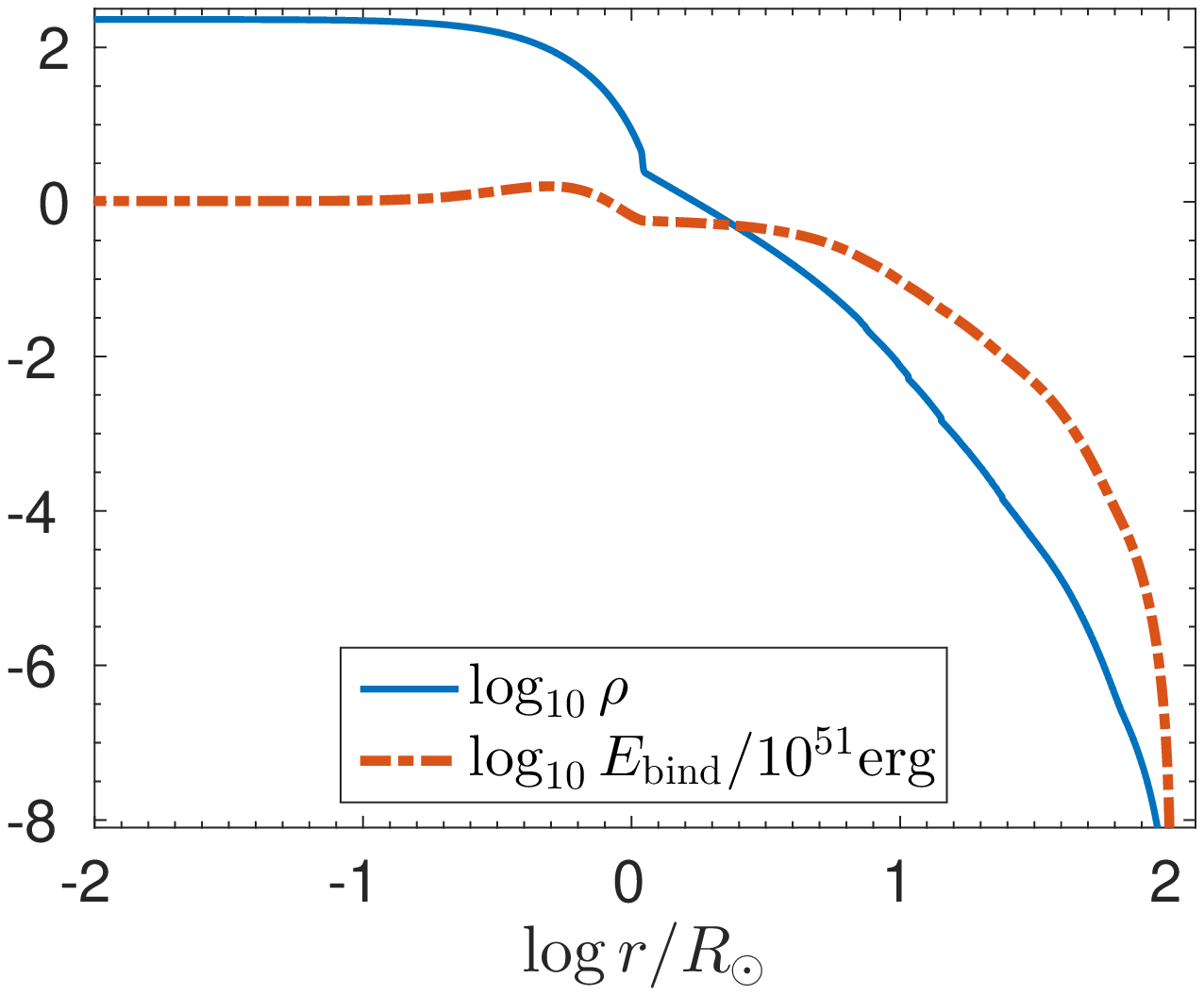}\\
\end{tabular}
\vskip +0.5 cm
\caption{\textit{Top:} Mass fraction of the predominant elements of a stellar model with an initial mass of $M_\mathrm{ZAMS}=80M_\odot$ at the time its radius is  $100 R_\odot$.
\textit{Middle:} Enclosed mass as function of radius.
\textit{Bottom:} Density profile at $R=100R_\odot$, as well as the binding energy at each mass coordinate, i.e., the amount of additional energy needed to unbind matter from the surface down to that point.
}
\label{fig:star}
\end{center}
\end{figure}

To obtain some representative values we evolve a non-rotating stellar model with a zero age main sequence mass of $M_\mathrm{ZAMS}=80M_\odot$ and metallicity of $Z=0.001$. We use the stellar evolution code Modules for Experiments in Stellar Astrophysics (MESA version 10108; \citealt{Paxton2011,Paxton2013,Paxton2015}). Mass loss is according to \cite{Vink2001}, with a reduction factor of $0.33$ (for a review on mass loss see \citealt{SmithMassLossReview}). For the low metallicity we take, this results in a small mass lost upon reaching the giant phase which interests us, $M_\mathrm{lost}\approx 1 M_\odot$. In Fig. \ref{fig:starhistory} we present the evolution with time of the stellar radius and total mass lost by winds, and in Fig. \ref{fig:star} we present the stellar model when the radius is $R_1=100 R_\odot$.
%

To have a rapid spiraling-in process already from the surface, the binary system should be unstable to the Darwin instability. For the model presented in Fig. \ref{fig:star}, the moment of inertia is $I_1=\eta M_{\rm env} R_1^2$ with $\eta=0.0072$, where $M_{\rm env} = 46.5 M_\odot$ is the envelope mass. For an earlier time when the radius is $50 R_\odot$ we find for the same model $\eta=0.0137$. We note that ours is just one model, and a more massive model might have a more massive envelope at the relevant stage. The condition of Darwin instability for a circular orbit (later we will argue that the orbit was actually somewhat eccentric) reads 
\begin{equation}
M_2 < \frac{3I_1}{a^2_{\rm D}} \simeq 1.5
\left( \frac{\eta}{0.01} \right) 
\left( \frac{M_{\rm env}}{50 M_\odot} \right) 
\left( \frac{a_{\rm D}}{R_1} \right)^{-2} M_\odot.
     \label{eq:m2}
\end{equation}
This shows that we require the binary companion to be a neutron star, rather than a black hole. 

The demand for a very low mass companion comes also from the finding that the polarization of iPTF14hls is very low, implying an almost spherical explosion \citep{Arcavietal2017}. Namely, the companion cannot deform much the giant envelope as it enters the common envelope phase. 

The binding energy of the hydrogen envelope in the model presented in Fig. \ref{fig:star} is $E_{\rm bind} = 5.8 \times 10^{50} \erg$,  much below the estimated kinetic energy of the CSM of iPTF14hls ($E_{\rm CSM} \approx 10^{52} \erg$). The presence of hydrogen in the envelope at explosion time implies that the pre-explosion energy deposition cannot be below the base of the hydrogen-rich mass. Namely, it cannot come from the core. A secondary star that launches jets inside the envelope can deposit energy inside the envelope. With an efficiency of $\epsilon \simeq 0.02$ of converting rest  energy of the accreted gas $M_{\rm acc} c^2$ (say a fraction of 0.1 of the accreted mass is launched at terminal velocity of $v_{\rm jet} = 0.55c$), the accreted mass should be  
\begin{equation}
M_{\rm acc} \simeq 0.3 \left( \frac{E_{\rm CSM}}{10^{52} \erg} \right) \left( \frac {\epsilon}{0.02} \right)^{-1}  M_\odot. 
     \label{eq:massacc}
\end{equation}
The neutron star can stay a neutron star after an accretion of such an amount of mass. Within 15 days the accretion rate will be $\approx 10 M_\odot \yr^{-1}$.  Neutrino cooling of the accreted mass allows such a high accretion rate (e.g., \citealt{HouckChevalier1991}).  

In the jet feedback mechanism the relation $E_{\rm bind} \ll E_{\rm CSM}$ implies that the feedback efficiency is low. The reason is that during the common envelope phase the jets eject envelope gas above and below the location of the compact companion that launches the jets, while the companion accretes mass mainly from envelope zones inner to its orbits. Namely, the jets do not directly remove gas from the reservoir of the accreted gas. This makes the negative feedback cycle inefficient, and explains why the total energy in the jets was much larger than the binding energy of the envelope. We note that the gravitational energy that was released by the neutron star during its spiral-in to an orbital separation of $\simeq 1 R_\odot$ is only $\approx 10^{50} \erg$ and does not add much to the energy of the ejected envelope. For the same reason the direct binary interaction does not cause much deviation from spherical symmetry.  

We adopt the jet-driven envelope ejection over the alternative scenario of the pair-instability pulsations mechanism. \cite{RakavyShaviv1967} were the first to find that very massive stars are expected to suffer pair instability that results from pressure drop due to electron-positron pair production. A later paper by \citealt{Barkatetal1967} studied the explosion itself (although it was submitted after the  paper by Rakavy \& Shaviv, it appeared first, and even today misleads researchers on the truly first paper on stellar pair instability). \cite{Arcavietal2017} consider the possibility that pair instability pulsations lead to the pre-explosion outbursts, but dismiss it for two reasons. (1) The first pair instability pulsation, that takes place in the stellar core, is expected to eject the entire hydrogen-rich envelope. (2) The pair instability is not expected to supply the energy of $E_{\rm CSM} \approx 10^{52} \erg$. We note that our examined $80 M_\odot$ stellar model might become pair-unstable, but only after expansion and formation of an oxygen core, while our proposed scenario occurs earlier (during expansion) and prevents a pair-instability explosion.

\section{THE EXPLOSION}
\label{sec:explosion}

There are two possible explanations for the final explosion of iPTF14hls. In the first the core of the giant star experienced a regular core collapse supernova. However, this cannot work with our proposed pre-exploison outburst scenario. If the companion is to spiral-in within less than 100 days (constraint by the non-detection of pre-explosion outburst 100 days before discovery), the giant could not be much larger than about $200 R_\odot$. But a star with a rich hydrogen envelope will reach larger radii before it explodes (see figure \ref{fig:starhistory}). As well, the slow evolution of the supernova suggests that it is not a regular core-collapse supernova. 
   
We instead follow \cite{Chevalier2012} and \cite{Papishetal2015} and attribute the explosion to the rapid accretion of core material onto the neutron star companion as the latter merges with the core. For that, \cite{Chevalier1996} and then \cite{Papishetal2015} in a stronger way consider impossible the formation of a Thorne--Zytkow object (TZO), where a neutron star settles at the center of the star \citep{ThorneZytkow1975}. 
Overall, the final explosion is powered by a neutron star  that accretes mass at a high rate from the core and launches jets that power the explosion. 
 
In the jet feedback explosion mechanism of most core-collapse supernovae the explosion energy is about several times the binding energy of the core (e.g., \citealt{PapishSoker2011}; review by \citealt{Soker2016Rev}). Simply, this is the energy that is required for the jittering jets \citep{PapishSoker2014} to expel the core and terminate accretion onto the newly born neutron star. When the jets are well collimated they are less efficient in removing the core material from the equatorial plane. The accretion process continues and the jets carry much more energy than the binding energy of the core. This inefficient jet feedback process leads to a super-energetic (and/or super-luminous) supernova \citep{Gilkisetal2016}.

As discussed in section \ref{sec:removal}, when the jets are launched by a neutron star that accretes mass from the envelope the feedback is inefficient. While the jets are launched perpendicular to the equatorial plane and remove envelope mass from those regions, the mass supply comes from inner envelope zones and through the equatorial plane. This inefficient negative feedback mechanism implies that in total the jets carry much more energy than the binding energy of the envelope they remove. As discussed by \cite{Papishetal2015}, the core is destructed and forms an accretion disk around the neutron star. The disk launches jets that further energize the expanding gas. Fallback material can prolong even more the accretion process and augment the explosion energy.  \cite{Arcavietal2017} find that at very late times in their observations the light curve declines as $t^{-5/3}$, supporting a powering by fallback material. Furthermore, instabilities in the accretion flow and interaction of jets with earlier ejecta might account for the light-curve variability and multiple peaks. 
 
An alternative explanation for the peaks is the collision of the ejecta with previously ejected mass. \citealt{Arcavietal2017} dismiss collision with circumstellar matter because they see no narrow lines. However, in our model the previously ejected gas moves at thousands of $\km \s^{-1}$, and we do not expect to see narrow lines. This surrounding gas moves only slightly slower that the photosphere, also at thousands of $\km \s^{-1}$. The collision occurs inside the broad-line forming region and hence does not dilute the lines.

The entire process from the onset of the common envelope phase to the final explosion is a continuous one. But there is a qualitative difference as the neutron star reaches the core because the time scales is much shorter now and the density of the core is much larger (see density profile in Fig. \ref{fig:star}). The spiraling-in process inside the envelope lasts for weeks. The outer parts that were ejected first reach a distance of $\approx 10^{15} \cm$ while the inner envelope region is ejected. 
The final spiraling-in process, from a radius of $\approx 5 R_\odot$ to the core, lasts for only about a day. The hydrogen-rich envelope that is ejected from this volume does not reach a large distance before the neutron star reaches the core, accretes mass within several hours and launches energetic jets that drive shock waves through the expanding dense envelope. This results in a supernova explosion. During most of the observation time the photosphere resides within the mass that was ejected in this last phase of neutron star interaction with the dense core. 

We term this entire explosion a \textit{common envelope jets supernova}. 

In the premise where all core collapse supernovae are driven by jets (e.g. \citealt{Papishetal2015a}), common envelope jets supernovae should qualitatively be similar to core collapse supernovae, and in the present case a type II-P supernova. 
   
\section{SUMMARY}
\label{sec:summary}

We proposed that both the pre-explosion mass ejection and the explosion of the type II-P SN iPTF14hls were powered by a neutron star that accreted mass while spiraling-in inside the envelope and the core of a giant massive star. We argue that this \textit{common envelope jets supernova} scenario can account for the basic properties of iPTF14hls. 

The process operates by a negative feedback mechanism, where the jets remove mass from the reservoir from which mass is accreted to power them. However, because the jets are launched perpendicular to the orbital plane while the accretion flow onto the neutron star is from inner zones of the giant and near the equatorial plane, the feedback cycle is inefficient. This explains  the kinetic energy of the outer gas that is responsible for some absorption lines being much larger than the binding energy of the envelope. 

When the core is destructed by the neutron star, the process is somewhat different as the core is expected to form an accretion disk around the neutron star, and the explosion energy is about the binding energy of the core. The explosion takes place when the neutron star is interacting with the helium rich core. However, the inner regions of the hydrogen-rich envelope are still with-in hundreds of solar radii, and the explosion shock catches up with those layer and makes the explosion a type II-P SN.   

We attribute the outburst of 1954, which is essentially an intermediate luminosity optical transient (ILOT), to an eccentic orbit of the neutron star before it entered the envelope. The event of high mass accretion rate at periastron passage, qualitatively similar to the binary model of the Great Eruption in Eta Carinae \citep{KashiSoker2010}, could have powered this ILOT event.  
 
An earlier ejection of mass can lead to the interaction of the SN ejecta with the earlier ejected mass, the CSM. \cite{AndrewsSmith2018} find evidence for an interaction with CSM expanding at $\approx 1000 \km \s^{-1}$ at late times and suggest that iPTF2014hls can be a regular CCSN interacting with the CSM. We see two problems with this scenario. (1) They do not account for the massive and fast ($\approx 6000 \km \s^{-1}$) absorbing material that \cite{Arcavietal2017} find. (2) Their scenario involves collision deep inside the ejecta. This has a low radiation efficiency, because part of the collision energy goes to accelerating the slowly expanding earlier ejecta, and because the long diffusion time of the X-ray photons out (otherwise X-ray emission will be strong). So even in their scenario the explosion energy is very large. We do agree with them that ejecta-CSM interaction does take place and does contribute, in particular at late times, but our aim is to account for the massive and fast absorbing gas that \cite{Arcavietal2017} claim for.

One challenge to the present model is the finding of \cite{Arcavietal2017} that the polarization is very low, implying a spherical explosion. We account for this as follows. The low mass companion, with a mass of $M_2 \simeq 0.02 M_1$, implies that it cannot deform much the envelope of the giant star. The jets were launched deep in the envelope and did not break out due to the fast orbital motion \citep{Papishetal2015}. Their energy deposition expelled the envelope and then the core to all directions. 

The evolution of the progenitor binary system leading to such a \textit{common envelope jets supernova} event is also of interest. One possible channel is through immense mass transfer from a primary of $M_\mathrm{ZAMS}\approx 50 M_\odot$ onto a secondary of $M_\mathrm{ZAMS}\approx 40 M_\odot$, with the primary then exploding as a type Ib or type Ic supernova (with a mass of few$ \times M_\odot$) leaving behind a neutron star. A $50M_\odot$ star will collapse before a $40M_\odot$ star leaves the main-sequence, but the mass transfer will alter the evolution time, so a detailed binary evolution model is required to study the path towards the scenario described in this work. Also, hydrodynamical simulations of the interaction between the neutron star and the envelope (and the neutron star and the core) are needed to further elucidate the scenario.

Additional study of our proposed \textit{common envelope jets supernova} scenario is required. At present we raise the prediction that the inner part of the ejecta might show bipolar structure due to the last jet-launching episodes.

\section*{Acknowledgments}

We thank Amit Kashi and Erez Michaely for helpful comments. This research was supported by the Israel Science Foundation and by the Pazi grant. A.G. is supported by the Blavatnik Family foundation.

\ifmnras
\bibliographystyle{mnras}

\label{lastpage}
\end{document}